# Gate-tunable ferromagnetism in a van der Waals magnetic semimetal

Hideki Matsuoka[1,3], Shun Kajihara[2,3], Yue Wang[2], Yoshihiro Iwasa[1,2]★ and Masaki Nakano[1,2]★

[1]*RIKEN Center for Emergent Matter Science (CEMS), Wako 351-0198, Japan.*

[2]*Quantum-Phase Electronics Center and Department of Applied Physics, the University of Tokyo, Tokyo 113-8656, Japan.*

[3]*These authors contributed equally: H. Matsuoka, S. Kajihara*

★e-mail: iwasa@ap.t.u-tokyo.ac.jp / nakano@ap.t.u-tokyo.ac.jp

**Magnetic semimetals form an attractive class of materials because of the non-trivial contributions of itinerant electrons to magnetism. Due to their relatively low-carrier-density nature, a doping level of those materials could be largely tuned by a gating technique. Here we demonstrate gate-tunable ferromagnetism in an emergent van der Waals magnetic semimetal $Cr_3Te_4$ based on an ion-gating technique. Upon doping electrons into the system, the Curie temperature ($T_C$) sharply increases, approaching near to room temperature, then decreases to some extent. Interestingly, this non-monotonous variation of $T_C$ accompanies the switching of the magnetic anisotropy. Furthermore, such evolutions of $T_C$ and anisotropy occur synchronously with the sigh changes of the ordinary and anomalous Hall effects. Those results clearly elucidate that the magnetism in $Cr_3Te_4$ should be governed by its semimetallic band nature, where the band crossing points play a crucial role both for the magneto-transport properties and magnetism itself.**

Research on itinerant magnetism has quite a long history starting from the antient period, but it continues to be a major topic in modern condensed-matter physics. Of particular importance is to understand an interplay between itinerant electrons and local magnetic moments, which provides an insight into the origin of magnetism. As for a material system, magnetic semimetals form an important branch, where the intersection of electronic bands with different spin and orbital characters plays an essential role for their non-trivial magneto-transport properties. The unique features of this system have been demonstrated in a variety of materials including magnetic topological insulators[1–4], magnetic Weyl semimetals[5–7], ferromagnetic/antiferromagnetic Kagome systems[7–9], and magnetic two-dimensional (2D) materials[10–13], where an interplay between the crossing points of the electronic bands and local magnetic moments leads to the unprecedented magneto-transport phenomena as represented by the (half-)quantization of the anomalous-Hall effect (AHE) or giant AH angle, which is associated with the emergence of the Berry curvature at the band-crossing points[14–17]. On the other hand, in those systems, the origin of magnetism has been of great interest as well. In particular, the unique carrier-density dependence of the Curie temperature ($T_C$) in magnetic topological insulators has been attributed to the non-trivial contributions of itinerant Dirac electrons to magnetism[4,18], which should be relevant to the energetic stability of the spontaneous magnetic ordering. These unique features could be generally observed in magnetic semimetals hosting both electron- and hole-like bands that are crossing near the Fermi level ($E_F$), but so far, such a discussion has been mostly limited to magnetic topological systems. To deepen our knowledge on the origin of magnetism in magnetic semimetals, it is highly desired to explore a new material system and examine the impact of a semimetallic band nature on the overall magnetic properties.

In this study, we focus on $Cr_3Te_4$ as an ideal material platform. As will be introduced later, $Cr_3Te_4$ is an emergent van der Waals (vdW) magnetic semimetal, whose magnetic properties are expected to be strongly coupled to its semimetallic band nature. To systematically control a doping level (*i.e.*, $E_F$) in $Cr_3Te_4$, we apply an electrical gating technique. The gate-controlled tuning of magnetism has been rather established approach, providing significant achievements both in magnetic semiconductors and even in magnetic metals[19–23], where $T_C$ could be largely enhanced by increasing the carrier density. However, in the case of magnetic semimetals, we could expect to have non-trivial contributions of itinerant electrons near the band-crossing points, which might provide an anomalous carrier-density dependence of their magnetism.

**Results**

**Basic properties of $Cr_3Te_4$ epitaxial thin films.**

$Cr_3Te_4$ is one of the self-intercalated phases of chromium telluride $(Cr_{1+\delta}Te_2)$[24], consisting of the host 2D $CrTe_2$ layers and the intercalated Cr layers that form the 1 × 2 superstructure within the vdW gap as illustrated in Figs. 1a and 1b. $Cr_{1+\delta}Te_2$ is a rather classical material system[25–27], but its relatively high $T_C$ has attracted considerable interest in the context of "2D magnet" toward potential 2D spintronics applications, and a growing number of papers has been published very recently including demonstrations of room-temperature 2D ferromagnetism[28–31]. On the other hand, recent fundamental studies on $Cr_{1+\delta}Te_2$ epitaxial thin films grown by molecular-beam epitaxy (MBE) have demonstrated their intriguing magnetic properties, which are even distinct from those of their bulk counterparts[32–35]. In this study, we started a series of the gating experiments with the metastable ferromagnetic phase of $Cr_3Te_4$ epitaxial thin films characterized with $T_C$ ~160 K and large out-of-plane magnetic anisotropy, which were prepared by MBE by following our previously-established growth process (see Methods)[34].

Figure 1c shows the magnetization versus temperature (*M-T*) curves of a typical sample taken with the out-of-plane (red) and in-plane (blue) magnetic fields. The *M-T* curve with the out-of-plane field showed significantly larger signals than that with the in-plane field, indicating its large out-of-plane magnetic anisotropy, which is one of the unique features of the metastable ferromagnetic phase of $Cr_3Te_4$ epitaxial thin films[34]. $T_C$ of this particular sample was determined to be 170 K, corresponding to the onset temperature of the *M-T* curve. Importantly, this onset of ferromagnetism could be well defined in the sheet resistance versus temperature ($R_s$-*T*) curve as a "kink" temperature (Fig. 1d), which enables us to track the variation of $T_C$ with increasing a doping level in the following gating experiments. Figures 1e and 1f illustrate a schematic band structure of $Cr_{1+\delta}Te_2$ epitaxial thin films as revealed by the angle-resolved photoemission spectroscopy measurements[35], which verified the existence of a characteristic semimetallic band region near $E_F$, where electron- and hole-like bands intersect with each other. The crossing points are accessible by a thermal-annealing treatment, which results in switching $T_C$ from 160 K to above-room temperature and magnetic anisotropy from the out-of-plane easy-axis type to the in-plane easy-plane type[32,33,35]. Those results imply that the carrier density should play an important role for the magnetism in $Cr_{1+\delta}Te_2$, although the effect of a thermal-annealing treatment remains unclear.

**Gate-tunable ferromagnetism in $Cr_3Te_4$.**

To tune a doping level largely across the semimetallic band region, we chose an *in-situ* Li intercalation technique, which has been widely used for controlling physical properties of a variety of layered materials through a reversible carrier doping process[22,36–38]. We fabricated ion-gating devices with a top-gate configuration (see the inset of Fig. 2a) by using a polymer electrolyte with $LiClO_4$. The details of the device fabrication process and the gating procedure are described in Methods. Figure 2a shows the $R_s$-$T$ curves of a typical device taken at different gate voltages ($V_G$), providing an overview of the gating effects in $Cr_3Te_4$. Above a threshold $V_G$, $R_s$ started to increase monotonously with increasing positive $V_G$, which corresponds to electron doping. This behavior is quite reasonable considering that $Cr_3Te_4$ is a hole-type conductor[34], and the increase of $R_s$ could be attributed to the depletion of hole carriers. Interestingly, we observed a significant modulation of the kink temperature above a threshold $V_G$, which appeared to be non-monotonous against $V_G$ and approaching near to room temperature at the intermediate regime ($V_G$ = 3.25 V). In the following, we categorize the obtained results into three regions, Phase I ($V_G$: 0 V ~ 3.1 V), Phase II ($V_G$: 3.2 V ~ 3.3 V), and Phase III ($V_G$: 3.4 V ~ 3.45 V), and discuss the magnetic properties at each phase based on the magneto-transport results. We note that the observed gating effects were highly reversible and reproducible among different devices (see Supplemental Information Sections A and B), suggesting that electron doping was induced by a reversible Li intercalation process rather than irreversible chemical reactions such as Te removal.

Figure 2b demonstrates the variation of the anti-symmetrized Hall resistance ($R_{yx}$) versus magnetic field curves with increasing a doping level taken at the lowest temperature ($T$ = 2 K) at each $V_G$. The magnetic fields were set to be the out-of-plane direction. $R_{yx}$ of a magnetic material is usually known to have two components, the ordinary Hall component proportional to the external magnetic fields and the AH component proportional to the magnetization. In Fig. 2b, $R_{yx}$ above 2 T was linear to the external fields for all $V_G$, which could be attributed to the ordinary Hall effect (OHE). With increasing $V_G$, OHE eventually showed a sign change from positive to negative as highlighted in the inset of Fig. 2b for two representative voltages, $V_G$ = 0 V and $V_G$ = 3.4 V, evidencing that the carrier type was changed from holes to electrons in $Cr_3Te_4$ upon electron doping. As will be explained later, the slope of OHE corresponding to the Hall coefficient ($R_H$) exhibited non-monotonous behavior against $V_G$ with a peak-like structure within Phase II, and the sign change occurred at the boundary between Phase II and Phase III. In addition, very interestingly, we verified that the sign of AHE was also switched from negative to positive exactly at a specific $V_G$ where $R_H$ recorded the maximum value in Phase II, suggesting their correspondence. Importantly, these sign changes of OHE and AHE occurred continuously without introducing an insulating phase at the intermediate regime as shown in Fig. 2a. This directly proves that a semimetallic band region hosting both electron- and hole-like bands illustrated in Fig. 1f should exist in the present sample as well, and that $E_F$ should pass through this region within the applied $V_G$ range. Moreover, the observed quite systematic and synchronous evolutions in OHE and AHE suggest that AHE should have the intrinsic origin associated with the band structure, and that there should be a singularity of the Berry curvature in the momentum space near $E_F$, most likely associated with the crossing points in the semimetallic band region.

Now we turn to focus on the shape of AHE. We found that the evolution of the shape of AHE occurred in two steps. At Phase I, a square-shaped AHE with a sizable hysteresis loop was observed, corresponding to the out-of-plane magnetic anisotropy. On the other hand, this hysteresis completely disappeared at Phase II, and again opened up at Phase III. The observed non-linear behavior without hysteresis at Phase II could be attributed to AHE with the in-plane magnetic anisotropy, where the AHE signal should increase gradually as the magnetization direction is changed from the in-plane direction at zero field to the out-of-plane direction at finite out-of-plane external fields. This series of changes in AHE across the three phases should therefore suggest that the magnetic anisotropy was changed from the out-of-plane type to the in-plane type, and finally switched back to the out-of-plane one. Such a sharp, complete, and consecutive switching of the magnetic anisotropy depending on a doping level has never been observed in any other magnetic materials thus far, highlighting one of the unique aspects of the current system. Figure 2c shows the corresponding evolution of the symmetrized magnetoresistance (MR), which is fully consistent with that of AHE: a butterfly-shaped MR and a square-shaped AHE were commonly observed both at Phase I and Phase III, while a negative, broad, and less-hysteretic MR emerged at Phase II, where AHE exhibited non-linear behavior without hysteresis as mentioned above.

The obtained experimental results support the idea that the magnetic anisotropy was switched two times across the three phases. In particular, a square-shaped AHE and a butterfly-shaped MR with clear hysteresis at Phase I and Phase III provide decisive evidence that those two phases host the out-of-plane magnetic anisotropy. Figures 3a and 3c summarize the AHE (red) and MR (blue) signals at Phase I and Phase III, respectively, showing a good correspondence with each other. The peak magnetic field of the MR curve should correspond to the coercive field ($H_c$), at which the magnetization changes its direction from one to the other along the external field direction. On the other hand, a signature of the in-plane anisotropy at Phase II characterized with a non-linear AHE and a broad MR with negligible hysteresis shown in Figs. 2b and 2c might be rather elusive. To address this issue, we performed additional MR measurements at Phase II with the *in-plane* magnetic fields. Figure 3b shows the corresponding MR data taken at $T = 2$ K with the in-plane fields at Phase II. It turned out that the shape of MR transformed from a broad, less-hysteretic one with the out-of-plane fields (Fig. 2c) to a butterfly-shaped, hysteretic one with the in-plane fields (Fig. 3b), decisively elucidating that Phase II has the in-plane magnetic anisotropy.

The red symbols in Figs. 4a, 4b, and 4c show the temperature dependences of $H_c$ at Phase I, Phase II, and Phase III, respectively, determined by the MR curves taken at each temperature (see Supplemental Information Section C). Note that the magnetic fields were set to be parallel to the easy axis directions (the same settings as in Fig. 3). Also shown by the blue symbols in Figs. 4a and 4c are those of $R_{AH}^{rem}$ at Phase I and Phase III, respectively, which is defined as the AH resistance ($R_{AH}$) value at zero field (see Fig. 3a) and deduced from the AHE data taken at each temperature (see Supplemental Information Section D). The $H_c$ and $R_{AH}^{rem}$ were verified to have the common onset temperature, corresponding to the onset of the spontaneous magnetization, which is nothing but $T_C$. Based on those magneto-transport results, we conclude that $T_C$ evolved from the lowest level at Phase I to the highest level at Phase II, and then to the intermediate level at Phase III with increasing $V_G$, demonstrating non-monotonous variation against a doping level. Owing to a good correspondence between $T_C$ determined by the magneto-transport measurements and the kink temperature in the $R_s$-$T$ curve as shown in Figs. 4d, 4e, and 4f, we could define $T_C$ at each $V_G$ from the $R_s$-$T$ data shown in Fig. 2a, and discuss the evolution of $T_C$ against a doping level more in detail.

**Magnetic phase diagram of Cr$_3$Te$_4$.**

Figure 5a summarizes the magnetic phase diagram of Cr$_3$Te$_4$ uncovered by the present ion-gating experiments, where $T_C$, $R_H$, and $R_{AH}$ at the saturated regime ($R_{AH}^{sat}$) are plotted as a function of $V_G$, highlighting the unique feature of this material system. Phase I is the initial state of the MBE-grown Cr$_3$Te$_4$, which is characterized with $T_C \sim 170$ K and large out-of-plane magnetic anisotropy. With increasing a doping level, the system enters into Phase II, where $T_C$ is largely enhanced near to room temperature. This phase change accompanies the switching of the magnetic anisotropy from the out-of-plane type to the in-plane one as schematically illustrated in the inset. On the other hand, $R_H$ increases as the depletion of hole carrier proceeds, and suddenly drops, showing a peak-like structure within Phase II, which accompanies the sign change of $R_{AH}^{sat}$. $R_H$ finally reaches to a negative value in Phase III, where $T_C$ is slightly decreased and the anisotropy returns back to the out-of-plane one. Interestingly, the large enhancement of $T_C$, the emergence of the in-plane magnetic anisotropy, the singular behavior in $R_H$, and the sign change of $R_{AH}^{sat}$ occur synchronously within Phase II, suggesting the same origin. Considering that $R_{AH}^{sat}$ changes its sign exactly when $R_H$ starts to decrease, a peak-like structure in $R_H$ is reflecting a semimetallic nature of this material system, and Phase II corresponds to the specific regime where $E_F$ is located nearly at the band crossing points. Such a non-monotonous variation in $T_C$ could not be explained by a simple Stoner model, where $T_C$ should be proportional to the density of states at $E_F$ as schematically drawn in Fig. 5b, which is totally different from our observations, where $T_C$ shows maximum near the band crossing points as illustrated in Fig. 5c, accompanying the two-step switching of the magnetic anisotropy. Interestingly, a similar carrier-density dependence of $T_C$ was observed in one of magnetic topological insulators, Mn-doped $(Bi,Sb)_2(Se,Te)_3$, where $T_C$ shows the maximum value exactly when $E_F$ is positioned at the Dirac point[18], implying that the observed anomalous carrier-density dependence of $T_C$ might capture a general feature of magnetism in magnetic semimetals.

**Discussion**

Here we discuss a possible origin of magnetism in the current system based on our observations. First, the strongly carrier-density-dependent nature indicates that the magnetism in $Cr_3Te_4$ should be governed by the itinerant carriers rather than the local exchange interaction. Second, a non-monotonous variation in $T_C$ does not agree with a simple Stoner model as mentioned above, which is clearly distinct from the previous gating results on a vdW magnetic metal, $Fe_3GeTe_2$ (Ref.22). Moreover, this unusual behavior is also unexpected within the framework of the conventional carrier-mediated RKKY mechanism[39–41], which supports a positive correlation between the ferromagnetic interaction (and thus $T_C$) and the carrier density at least in the low-carrier-density regime as widely demonstrated in diluted magnetic semiconductors[19-21]. On the other hand, for magnetic semimetals hosting a band crossing point in the momentum space, the valence-electron-mediated Van Vleck mechanism should have a large contribution[42], where an energy gain associated with the generation of the magnetic susceptibility due to the band hybridization plays an essential role as often discussed in magnetic topological insulators[43,44]. Given that the effect of the band mixing should be maximized when $E_F$ is positioned exactly at the band crossing points, the magnetism associated with the Van Vleck mechanism should be highly sensitive to $E_F$, and it should be weakened when $E_F$ is moved far away from the crossing points. In addition, considering that the band hybridization naturally generates orbital angular momentum in the momentum space, the magnetic anisotropy could be also modulated near the crossing points, which is exactly what we observed in our experiments. Taken together, we consider that the large enhancement of $T_C$ and the sudden change in the magnetic anisotropy at Phase II should be associated with the activation of the Van Vleck mechanism near the band crossing points. Considering that $T_C$ should be determined as the sum of the exchange interactions in the system, the large enhancement of $T_C$ at Phase II is somewhat within our expectation because the Van Vleck mechanism should purely "add" an extra contribution to another mechanism most likely the RKKY mechanism that governs the magnetism in $Cr_3Te_4$ in the entire regime even far away from the crossing points.

In summary, the present study demonstrates the successful modulation of the magnetic properties of an emergent vdW magnetic semimetal $Cr_3Te_4$ by an ion-gating technique, where the fundamental properties of a ferromagnet, $T_C$ and magnetic anisotropy, are largely tuned by $V_G$. In addition, based on the systematic and synchronous variations of the magnetic properties and the transport coefficients including OHE and AHE, we could get an insight into the origin of magnetism in this material system, where the non-trivial contributions of itinerant electrons near the crossing points to magnetism should play a crucial role. The suggested scenario could be generally applicable to other magnetic semimetals, providing a new perspective on designing a next generation 2D magnet with high-enough $T_C$ and large-enough magnetic anisotropy that are definitely required for future ultracompact nanoscale 2D spintronics applications.

**Online content**

Any methods, additional references, Nature Research reporting summaries, source data, statements of data availability and associated accession codes are available at ---.

**References**

1. Hasan, M. Z. & Kane, C. L. Colloquium: Topological insulators. *Rev. Mod. Phys.* **82**, 3045–3067 (2010).

2. Qi, X. L. & Zhang, S. C. Topological insulators and superconductors. *Rev. Mod. Phys.* **83**, 1057–1110 (2011).

3. Chang, C. Z. *et al.* Experimental observation of the quantum anomalous Hall effect in a magnetic topological insulator. *Science (1979)* **340**, 167–170 (2013).

4. Tokura, Y., Yasuda, K. & Tsukazaki, A. Magnetic topological insulators. *Nat. Rev. Phys.* **1**, 126–143 (2019).

5. Wan, X., Turner, A. M., Vishwanath, A. & Savrasov, S. Y. Topological semimetal and Fermi-arc surface states in the electronic structure of pyrochlore iridates. *Phys. Rev. B* **83**, 205101 (2011).

6. Xu, G., Weng, H., Wang, Z., Dai, X. & Fang, Z. Chern semimetal and the quantized anomalous Hall effect in $HgCr_2Se_4$. *Phys. Rev. Lett.* **107**, 186806 (2011).

7. Liu, E. *et al.* Giant anomalous Hall effect in a ferromagnetic kagome-lattice semimetal. *Nat. Phys.* **14**, 1125–1131 (2018).

8. Nakatsuji, S., Kiyohara, N. & Higo, T. Large anomalous Hall effect in a non-collinear antiferromagnet at room temperature. *Nature* **527**, 212–215 (2015).

9. Ye, L. *et al.* Massive Dirac fermions in a ferromagnetic kagome metal. *Nature* **555**, 638–642 (2018).

10. Lu, X. *et al.* Superconductors, orbital magnets and correlated states in magic-angle

bilayer graphene. *Nature* **574**, 653–657 (2019).

11. Sharpe, A. L. *et al.* Emergent ferromagnetism near three-quarters filling in twisted bilayer graphene. *Science (1979)* **365**, 605–608 (2019).

12. Serlin, M. *et al.* Intrinsic quantized anomalous Hall effect in a moiré heterostructure. *Science (1979)* **367**, 900–903 (2020).

13. Deng, Y. *et al.* Quantum anomalous Hall effect in intrinsic magnetic topological insulator $MnBi_2Te_4$. *Science (1979)* **367**, 895–900 (2020).

14. Haldane, F. D. M. Model for a Quantum Hall Effect without Landau Levels: Condensed-Matter Realization of the 'Parity Anomaly'. *Phys. Rev. Lett* **61**, 2015–2018 (1988).

15. Ohgushi, K., Murakami, S. & Nagaosa, N. Spin anisotropy and quantum Hall effect in the *kagomé* lattice: Chiral spin state based on a ferromagnet. *Phys. Rev. B* **62**, R6067 (2000).

16. Fang, Z. *et al.* The Anomalous Hall Effect and Magnetic Monopoles in Momentum Space. *Science (1979)* **302**, 92–96 (2003).

17. Xiao, D., Chang, M. C. & Niu, Q. Berry phase effects on electronic properties. *Rev. Mod. Phys.* **82**, 1959–2007 (2010).

18. Checkelsky, J. G., Ye, J., Onose, Y., Iwasa, Y. & Tokura, Y. Dirac-fermion-mediated ferromagnetism in a topological insulator. *Nat. Phys.* **8**, 729–733 (2012).

19. Ohno, H. *et al.* Electric-field control of ferromagnetism. *Nature* **408**, 944–946 (2000).

20. Yamada, Y. *et al.* Electrically induced ferromagnetism at room temperature in cobalt-doped titanium dioxide. *Science (1979)* **332**, 1065–1067 (2011).

21. Matsukura, F., Tokura, Y. & Ohno, H. Control of magnetism by electric fields. *Nat. Nanotechnol.* **10**, 209–220 (2015).

22. Deng, Y. *et al.* Gate-tunable room-temperature ferromagnetism in two-dimensional $Fe_3GeTe_2$. *Nature* **563**, 94–99 (2018).

23. Verzhbitskiy, I. A. *et al.* Controlling the magnetic anisotropy in $Cr_2Ge_2Te_6$ by electrostatic gating. *Nat. Electron.* **3**, 460–465 (2020).

24. Ipser, H., Komarek, K. & Klepp, K. O. Transition metal-chalcogen systems viii: The Cr-Te phase diagram. *J. Less-Common Met.* **92**, 265–282 (1983).

25. Lotgering, F. K. & Gorter, E. W. Solid solutions between ferromagnetic and antiferromagnetic compounds with NiAs structure. *J. Phys. Chem. Solids* **3**, 238–249 (1957).

26. Yamaguchi, M. & Hashimoto, T. Magnetic properties of $Cr_3T_4$ in ferromagnetic region. *J Phys. Soc. Japan* **32**, 635–638 (1972).

27. Freitas, D. C. *et al.* Ferromagnetism in layered metastable 1*T*-$CrTe_2$. *J. Phys. Condens. Matter* **27**, (2015).

28. Sun, X. *et al.* Room temperature ferromagnetism in ultra-thin van der Waals crystals of 1T-$CrTe_2$. *Nano Res.* **13**, 3358–3363 (2020).

29. Zhang, X. *et al.* Room-temperature intrinsic ferromagnetism in epitaxial $CrTe_2$ ultrathin films. *Nat. Commun.* **12**, 2492 (2021).

30. Chua, R., Zhou, J. & Yu, X. Room temperature ferromagnetism of monolayer chromium telluride with perpendicular magnetic anisotropy. *Adv. Mater.* **33**, 2103360 (2021).

31. Wu, H. *et al.* Strong intrinsic room-temperature ferromagnetism in freestanding non-van der Waals ultrathin 2D crystals. *Nat. Commun.* **12**, 5688 (2021).

32. Fujisawa, Y. *et al.* Tailoring magnetism in self-intercalated $Cr_{1+\delta}Te_2$ epitaxial films. *Phys. Rev. Mater.* **4**, 114001 (2020).

33. Lasek, K. *et al.* Van der Waals epitaxy growth of 2D ferromagnetic $Cr_{(1+\delta)}Te_2$

34. Wang, Y. *et al.* Layer-Number-Independent Two-Dimensional Ferromagnetism in $Cr_3Te_4$. *Nano. Lett.* **22**, 9964–9971 (2022).

35. Fujisawa, Y. *et al.* Widely Tunable Berry Curvature in the Magnetic Semimetal $Cr_{1+\delta}Te_2$. *Adv. Mater.* **5**, 2207121 (2023).

36. Yu, Y. *et al.* Gate-tunable phase transitions in thin flakes of 1T-$TaS_2$. *Nat. Nanotechnol.* **10**, 270–276 (2015).

37. Nakagawa, Y. *et al.* Gate-controlled BCS-BEC crossover in a two-dimensional superconductor. *Science (1979)* **372**, 190–195 (2021).

38. Tang, M. *et al.* Continuous manipulation of magnetic anisotropy in a van der Waals ferromagnet via electrical gating. *Nat. Electron.* **6**, 28–36 (2023).

39. Ruderman, M. A. & Kittel, C. Indirect Exchange Coupling of Nuclear Magnetic Moments by Conduction Electrons. *Phys. Rev.* **96**, 99–102 (1954).

40. Kasuya, T. A Theory of Metallic Ferro- and Antiferromagnetism on Zener's Model. *Progress of Theoretical Physics* **16**, 45–57 (1956).

41. Yoshida, K. Magnetic Proyerties of Cu-Mn Alloys. *Phys. Rev.* **106**, 893–898 (1957).

42. van Vleck, J, H. *The Theory of Electronic and Magnetic Susceptibilities*. (Oxford: Clarendon Press, 1932).

43. Cooper, L. N. *et al.* Quantized Anomalous Hall Effect in Magnetic Topological Insulators. *Science (1979)* **61**, 61–64 (2010).

44. Núñez, A. S. & Fernández-Rossier, J. Colossal anisotropy in diluted magnetic topological insulators. *Solid State Commun.* **152**, 403–406 (2012).

45. Momma, K. & Izumi, F. VESTA 3 for three-dimensional visualization of crystal, volumetric and morphology data. *J. Appl. Cryst.* **44**, 1272–1276 (2011).

nanolayers with concentration-tunable magnetic anisotropy. *Appl. Phys. Rev.* **9**, 011409 (2022).

**Acknowledgments**

We are grateful to Y. Okada, K. Ishizaka, T. Nomoto, M. Hirayama, R. Arita, and A. Tsukazaki for valuable discussions, and to B. K. Saika and K. Endo for their experimental help. This work was supported by Grants-in-Aid for Scientific Research (Grant Nos. 19H05602, 19H02593, 22H01949, 21K13888) and A3 Foresight Program from the Japan Society for the Promotion of Science (JSPS), and by PRESTO (Grant No. JPMJPR20AC) from Japan Science and Technology Agency. This study was carried out by the joint research of the Cryogenic Research Center, the University of Tokyo.

**Author contributions**

Y.W. and S.K. grew and characterized the samples, performed magnetization and transport measurements. S.K. and H.M. performed the gating experiments and analyzed the data. H.M., Y.I., and M.N. supervised this study. H.M., Y.I., and M.N. wrote the manuscript. All the authors discussed the results and commented on the manuscript.

**Competing interests**

The authors declare no competing interests.

**Additional Information**

**Reprints and permissions information** is available at ---.

**Correspondence and requests for materials** should be addressed to Y.I. or M.N.

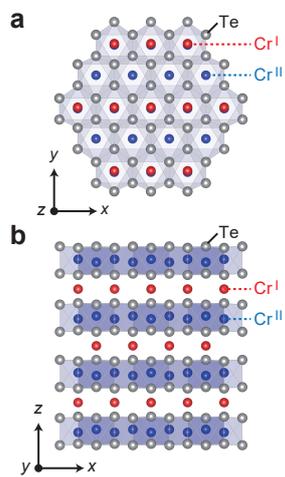
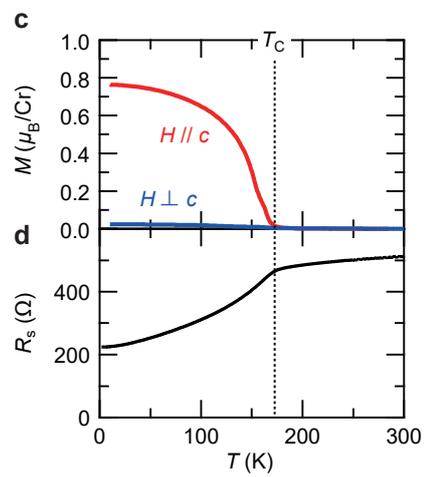
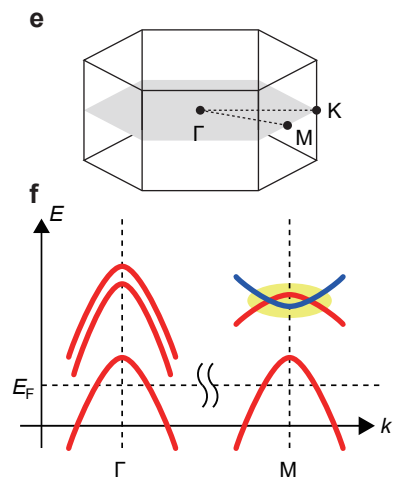

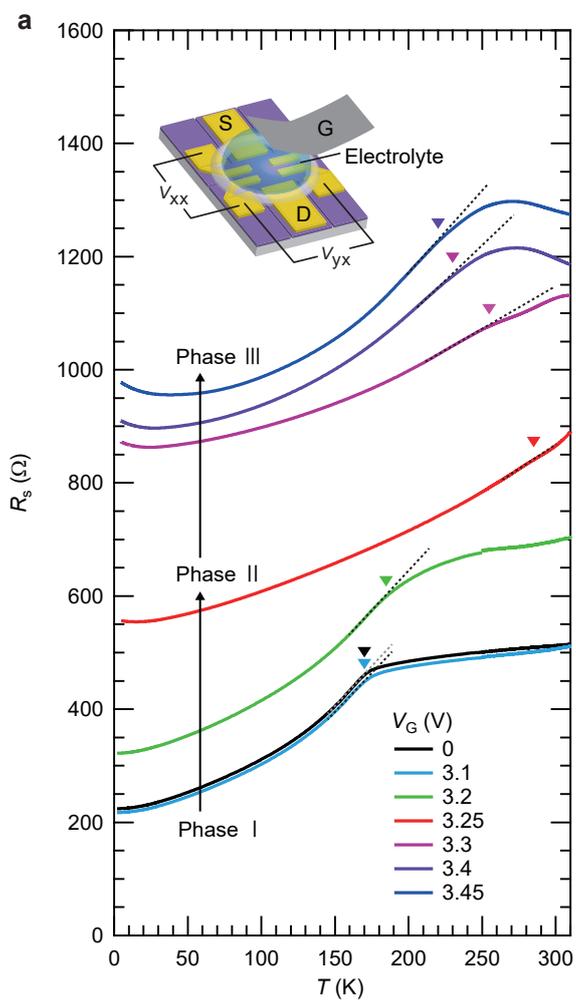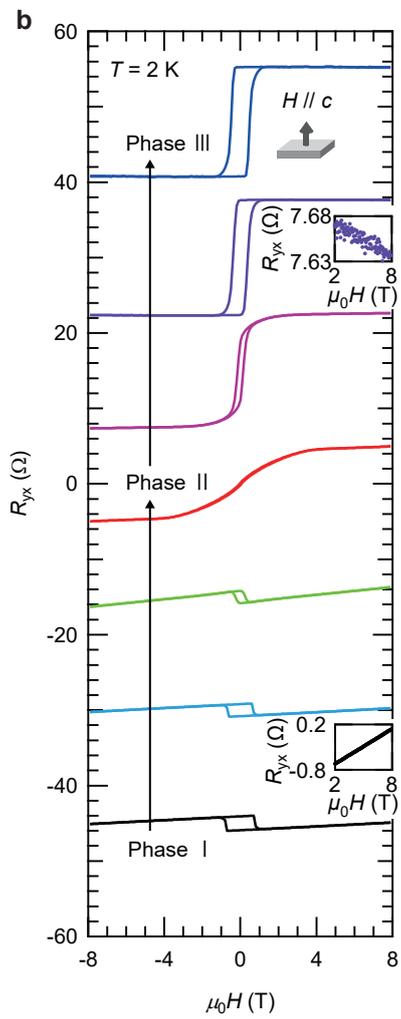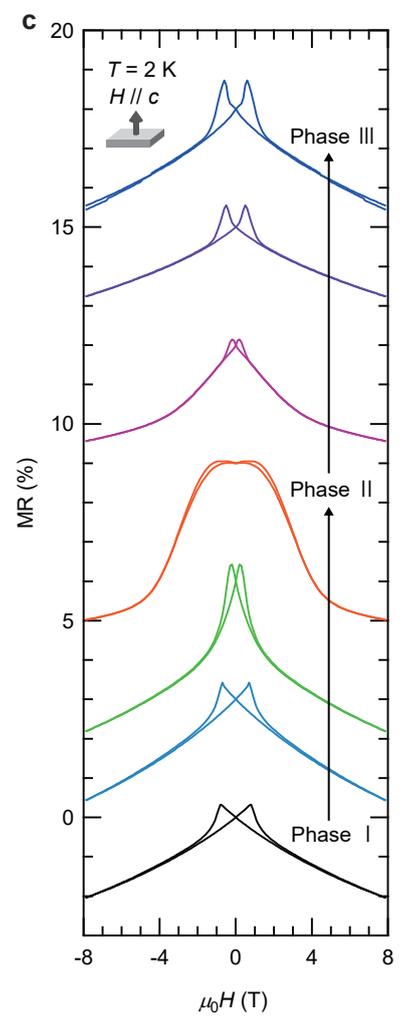

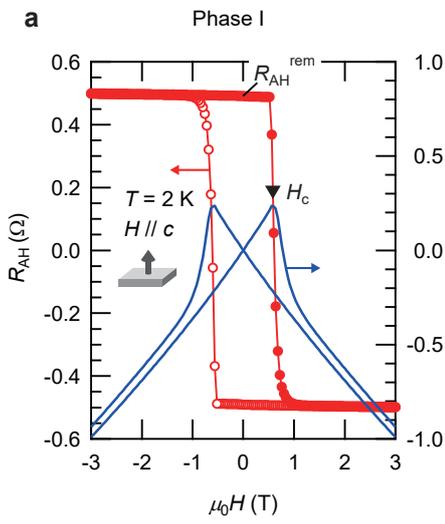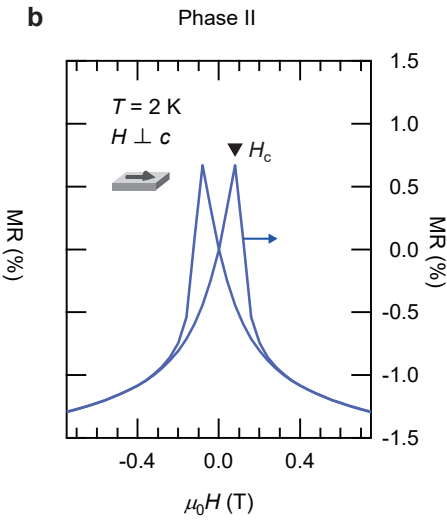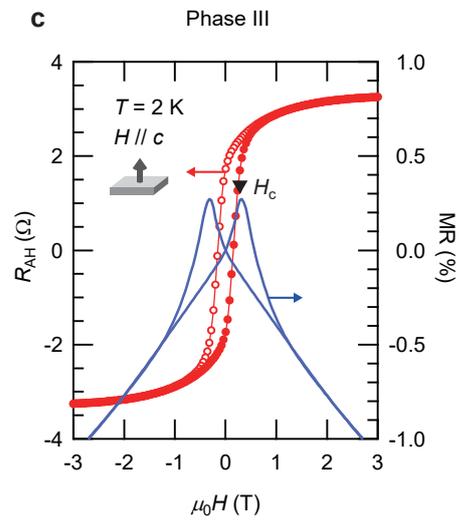

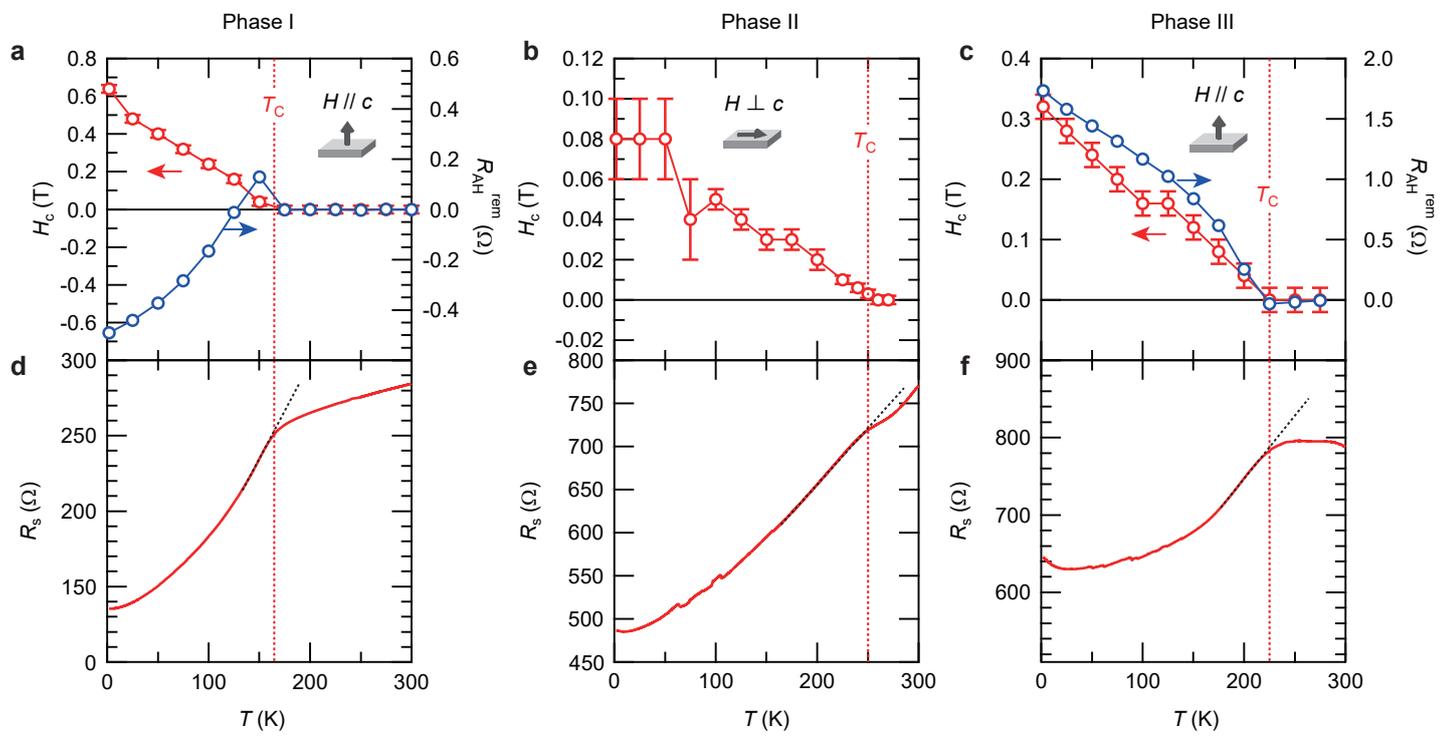

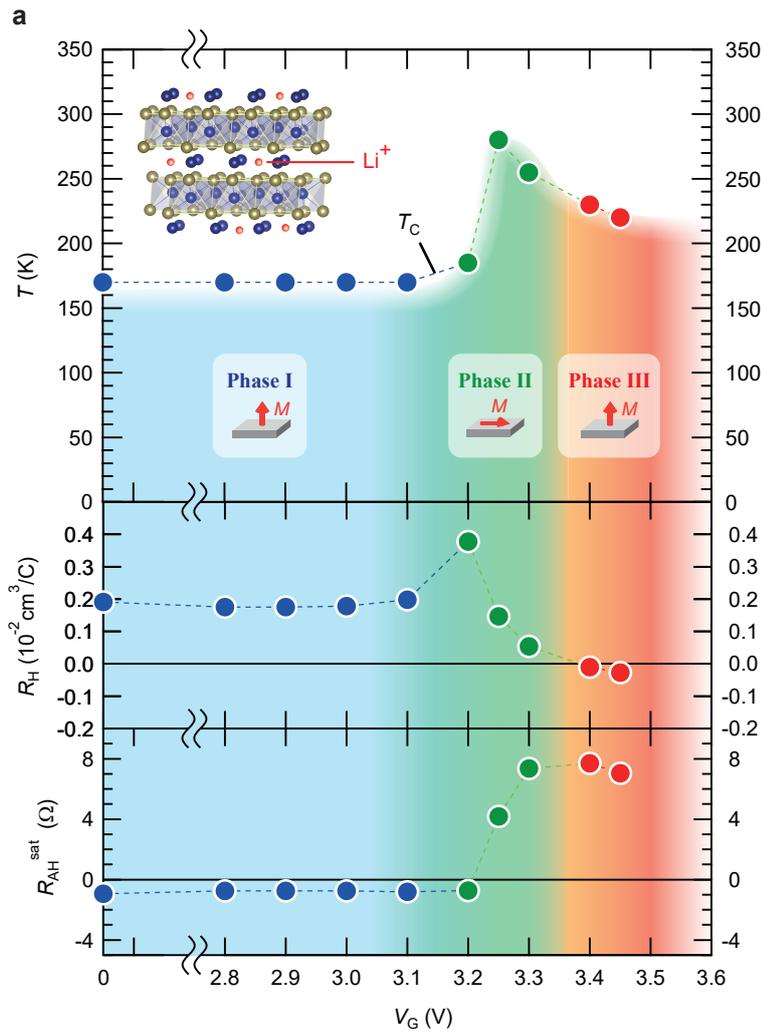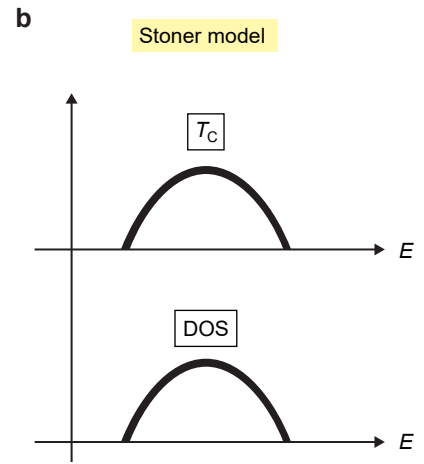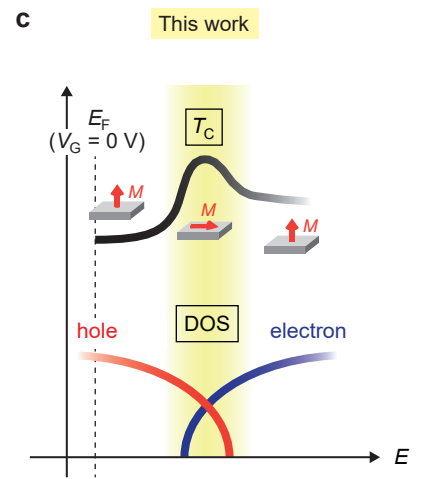

**Figure legends**

**Fig. 1 | Basic properties of $Cr_3Te_4$ epitaxial thin films.**

**a,b,** Schematic (**a**) top and (**b**) side views of the crystal structure of $Cr_3Te_4$ drawn by VESTA[45]. **c,** The magnetization versus temperature (*M-T*) curves of a typical $Cr_3Te_4$ epitaxial film taken in the field-cooling conditions with the out-of-plane (*H // c*, red) and in-plane (*H* ⊥ *c*, blue) magnetic fields. The magnitude of the applied field was $\mu_0H$ = 10 Oe. The layer number of this particular sample was 50 L, which is defined as the number of the host $CrTe_2$ layer. **d,** The corresponding sheet resistance versus temperature ($R_s$-*T*) curve of the same sample taken at zero field. The vertical dashed line corresponds to the Curie temperature ($T_C$), which is defined as the onset temperature of the spontaneous magnetization. Note that $T_C$ could be well defined in the $R_s$-*T* curve as a "kink" temperature, which is used to probe $T_C$ in the gating experiments. **e,f,** Schematic (**e**) first Brillouin zone and (**f**) band structure of $Cr_{1+\delta}Te_2$ as revealed by the angle-resolved photoemission spectroscopy measurements[35], where electron- and hole-like bands intersect with each other to form a semimetallic band region near the Fermi level ($E_F$) at the M point in the momentum space (yellow-colored area).

**Fig. 2 | Gate-tunable ferromagnetism in $Cr_3Te_4$.**

**a,** The $R_s$-*T* curves of a typical device (Device A) taken at different gate voltages ($V_G$). The inset shows a schematic device structure with a top-gate configuration. Phase I, Phase II, and Phase III are defined as the the $V_G$ regions of 0 ~ 3.1 V, 3.2 V ~ 3.3 V, and 3.4 V ~ 3.45 V, respectively. The inverse triangles correspond to $T_C$, which is defined from the kink temperature in the $R_s$-*T* curve. **b,c,** The magnetic-field dependences of (**b**) the anti-symmetrized Hall resistance ($R_{yx}$) and (**c**) the symmetrized magnetoresistance (MR) of Device A taken at *T* = 2 K at each $V_G$. The magnetic fields were set to be the out-of-plane directions. The inset two panels in **b** are the magnified views of $R_{yx}$ in the field range above 2 T for two representative voltages, $V_G = 0$ V and $V_G = 3.4$ V, demonstrating the sign change of the ordinary Hall effect (OHE) at a highly electron-doped regime. The MR is defined as $[R_{xx}(\mu_0 H) - R_{xx}(\mu_0 H = 0 \text{ T})]/R_{xx}(\mu_0 H = 0 \text{ T})$, where $R_{xx}$ corresponds to the symmetrized longitudinal resistance. All the $R_{yx}$ and MR data are vertically shifted for clarity except for the data in the inset two panels in **b**.

**Fig. 3 | Evaluation of the magnetic anisotropy.**

**a,b,c,** The magnetic-field dependences of $R_{AH}$ (red) and MR (blue) taken at $T = 2$ K at (**a**) Phase I (Device B, $V_G = 0$ V), (**b**) Phase II (Device C, $V_G = 3.4$ V), and (**c**) Phase III (Device B, $V_G = 3.4$ V). The magnetic fields were aligned to be parallel to the easy axis directions. $R_{AH}$ stands for the AH resistance, which is calculated from $R_{yx}$ by subtraction of the OH component. Note that Phase II does not have the $R_{AH}$ data, which was not measured because no AHE signal is expected for the in-plane fields. $R_{AH}^{rem}$ corresponds to $R_{AH}$ at zero field (see **a**). $H_c$ corresponds to the coercive field.

**Fig. 4 | Determination of $T_C$.**

**a,b,c,** The temperature dependences of $H_c$ (red) and $R_{AH}^{rem}$ (blue) at (**a**) Phase I (Device B, $V_G = 0$ V), (**b**) Phase II (Device C, $V_G = 3.4$ V), and (**c**) Phase III (Device B, $V_G = 3.4$ V). The magnetic fields were aligned to be parallel to the easy axis directions, the same settings as in Fig. 3. All the MR curves used to determine $H_c$ are shown in Supplemental Information Section C. The AHE data used to deduce $R_{AH}^{rem}$ are shown in Supplemental Information Section D. **d,e,f,** The corresponding $R_s$-$T$ curves taken at (**d**) Phase I (Device B, $V_G = 0$ V), (**e**) Phase II (Device C, $V_G = 3.4$ V), and (**f**) Phase III (Device B, $V_G = 3.4$ V). The red dotted lines represent the kink temperatures defined in the $R_s$-$T$ curves, exactly matching to the onset temperatures of $H_c$ and $R_{AH}^{rem}$.

**Fig. 5 | Magnetic phase diagram of $Cr_3Te_4$.**

**a,** The evolutions of $T_C$, the Hall coefficient ($R_H$), and $R_{AH}$ at the saturated regime ($R_{AH}^{sat}$) as a function of $V_G$. The blue, green, and red regions are corresponding to the three distinct phases, Phase I, Phase II, and Phase III, respectively. All the values plotted in this phase diagram were obtained from the data on Device A shown in Fig. 2. The inset shows a schematic of Li-intercalated $Cr_3Te_4$. **b,c,** Schematic illustrations of the correspondence between $T_C$ and the density of states (DOS) for (**b**) a simple Stoner model and (**c**) this work.

**Methods**

**Thin film growth and characterizations.**

$Cr_3Te_4$ epitaxial thin films were grown on sapphire (001) substrates by MBE (EIKO Engineering) by following our previously-established growth process[34]. The growth temperature was set to 450 °C. During the growth, Cr was supplied by an electron-beam evaporator with the evaporation rate below 0.05 Å/s, while Te was supplied by a Knudsen cell with the rate of ~ 3 Å/s throughout the growth process. All the films were annealed at 450 °C for 60 minutes after the growth with the same Te flux to improve the crystalline quality. The crystallinities of the obtained films were characterized by a reflection high-energy electron diffraction system (EIKO Engineering) and by a four-circle X-ray diffractometer (PANalytical, Empyrean). Magnetization measurements were performed by Magnetic Property Measurement System (Quantum Design, MPMS). We note that the structural phase of our samples was identified as $Cr_3Te_4$ by structural, magnetic, and transport characterizations in the previous study[34].

**Device fabrication.**

The ion-gating devices used in this study were fabricated by a standard process without lithography as follows. 10 nm-thick Ni and 100 nm-thick Au were deposited onto the films as source/drain electrodes and voltage probes by an electron-beam evaporator through a metal mask with a Hall-bar pattern, followed by mechanical scratching between each electrode to isolate voltage probes and define a channel size to be 100 μm in length and 300 μm in width. Subsequently, a Pt plate was placed above a channel region as a gate electrode, and a droplet of a polymer electrolyte was inserted in between a channel and a Pt plate just before the measurements. A polymer electrolyte was made of polyethylene glycol (PEG) ($M_w$ = 600, Wako) and $LiClO_4$ (Sigma Aldrich), which was heated at 80°C under vacuum before use. The ratio of Li to O in the PEG was set to 1:20.

**Transport measurements.**

The electrical transport properties of the ion-gating devices were characterized by combination of a source-measure unit (Agilent Technologies, B2912A) and voltmeters (Keithley Instruments, 2182A) at different temperatures and magnetic fields under different $V_G$ in Physical Property Measurement System (Quantum Design, PPMS). $V_G$ was applied at $T$ = 330 K under high vacuum condition (< $10^{-4}$ Torr). After cooling the system down to $T$ = 250 K, the sample chamber was purged with helium gas to improve the stability of temperature.